\documentclass[lettersize, journal]{IEEEtran}
\usepackage{lineno,hyperref}
\usepackage{amsmath}
\usepackage[ruled, lined, longend, linesnumbered]{algorithm2e}
\usepackage[square,numbers]{natbib}
\usepackage{amsmath, amsthm, amssymb}

\usepackage{amsthm}
\usepackage{placeins}
\usepackage{csquotes}
\usepackage{url}
\usepackage{multirow}
\usepackage{amssymb}
\usepackage{graphics}
\usepackage[utf8]{inputenc}
\usepackage[T1]{fontenc}
\usepackage{subcaption}
\usepackage{caption}
\usepackage{algpseudocode}
\usepackage{amsfonts}
\usepackage{filecontents}
\usepackage{mathtools}
\usepackage{booktabs}
\usepackage{enumitem}
\usepackage{epstopdf}
\setlist[itemize]{leftmargin=*}

\usepackage{xcolor}
\definecolor{rv1}{rgb}{1.0, 0.44, 0.37}
\definecolor{rv2}{rgb}{0.4, 1.0, 0.0}
\definecolor{rv3}{rgb}{0.0, 0.75, 1.0}
\definecolor{rvt}{rgb}{0.75, 0.75, 0.75}
\usepackage{soul}

\newtheoremstyle{exampstyle}
{7pt} 
{7pt} 
{\itshape} 
{} 
{\bfseries} 
{.} 
{.5em} 
{} 
\theoremstyle{exampstyle}

\algrenewcommand\algorithmicindent{0.5em}%
\ifCLASSINFOpdf
\graphicspath{{Figures/}}
\else
\fi
\newsavebox\mybox

\begin{document}
	\title{Polarforming-Enabled Power-Splitting SWIPT:\\
		A GNN-Based Optimization Approach}
	\author{\IEEEauthorblockN{~Hamed~Aghaei-Karkaj,~Kamran~Ebrahimi,~Zahra~Mehrzad,~\\Mohammad~Robat~Mili, and Symeon~Chatzinotas,~\textit{Fellow,~IEEE}
		}
		\thanks{
			H. Aghaei-Karkaj, Z. Mehrzad, and M. Robat Mili are with the Pasargad Institute for Advanced Innovative Solutions (PIAIS), Tehran, Iran, (e-mail: hamedaghaeikj@gmail.com, \{zahra.mehrzad, mohammad.robatmili\}@piais.ir).	
			K. Ebrahimi is with the Amirkabir University of Technology, Tehran, Iran, (e-mail: kamranebrahimi@aut.ac.ir). S. Chatzinotas is with the Interdisciplinary Centre for Security, Reliability and Trust (SnT), University of Luxembourg, L-1855 Luxembourg City, Luxembourg (e-mail: symeon.chatzinotas@uni.lu).
	}
}
	\maketitle
\begin{abstract}
Simultaneous wireless information and power transfer (SWIPT) is a critical technology for the future of the Internet of Things (IoT). However, ensuring a stable power supply in such networks remains a significant challenge. This work introduces dynamic polarization control as an additional degree of freedom (DoF) in SWIPT systems. We propose a system where both the base station (BS) and the users can adjust their antenna polarization, a technique known as polarforming. In addition, each user device is capable of splitting the incident signal to perform simultaneous information decoding (ID) and energy harvesting (EH). The resulting non-convex optimization problem, which involves strongly coupled variables, is addressed using a graph neural network (GNN) that predicts the beamforming, polarization, and power-splitting variables. Simulation results demonstrate that the proposed GNN-based dynamic polarforming optimization significantly outperforms fixed-polarization schemes, particularly under imperfect channel state information (CSI). Moreover, joint polarforming and GNN-based optimization maintains robust SWIPT performance under both polarization mismatch and imperfect CSI.
\end{abstract}
\begin{IEEEkeywords}
	Polarforming, power-splitting, SWIPT, graph neural network.
\end{IEEEkeywords}
\vspace{-1em}
\section{Introduction}
\IEEEPARstart{W}{ith} the rapid development of Internet of Things (IoT) applications in industry and environmental monitoring, the concept of simultaneous wireless information and power transfer (SWIPT) has emerged as a key enabling technology. However, ensuring a reliable power supply in these systems has become a significant challenge, particularly in remote environments. Moreover, conventional batteries suffer from several inherent limitations, including the need for maintenance requirements, replacement logistics, and restricted operational lifetimes in many practical settings such as agriculture, healthcare, and the environment \cite{haddad2026machine}.

Additionally, IoT systems are increasingly deployed in crowded urban environments and large-scale industrial settings, where there is a growing demand for high reliability, high data rates, and low energy consumption. These requirements necessitate the consideration of additional degrees of freedom (DoFs) in the design of such systems \cite{ding2025secure}. Recently, several other key technologies have been proposed to increase the DoFs. In \cite{ 11061790, ataeebojd2026energy}, metasurface-based approaches have been introduced to enhance system performance in terms of data rate, ultra-reliable low-latency communication (URLLC), and energy efficiency.
However, with metasurface-based technologies reaching their theoretical efficiency limits, considerable research interest has shifted toward reconfigurable antenna architectures, including movable antennas, pinching antenna systems, and fluid antenna systems \cite{ding2025energy, tegos2025minimum, liang2026rate}. Nonetheless, even with these advances in technology, the demand for high-capacity wireless links remains out of reach.

Research on wireless systems has predominantly been restricted to fixed linear and circular polarization configurations. Optimizing polarization as an additional design parameter for wireless communication systems offers the prospect of unlocking new DoFs, creating opportunities for simultaneously improving reliability, capacity, power efficiency, and security \cite{ding2026polarforming}. In addition, cost-efficient methods have been introduced to dynamically control the polarization of antennas. This approach employs a phase shifter to adjust the phase shifts of the antenna elements \cite{ding2025secure}.

Conventional wireless communication systems generally employ antennas with fixed polarization, such as linear and circular polarization, and thus cannot dynamically adapt their polarization states to changing propagation conditions. To overcome this limitation, polarforming has recently emerged as a promising technique that enables dynamic control of antenna polarization, thereby exploiting polarization as an additional design DoF \cite{ding2026polarforming}. In a phase-shifter-based implementation, each polarization-reconfigurable antenna comprises two co-located, orthogonally polarized antenna elements (vertical and horizontal), whose relative excitation phase is adjusted through a phase shifter to dynamically modify the resulting polarization state \cite{ding2025secure}. This capability enables adaptive polarization matching or mismatching with the received electromagnetic (EM) waves, offering opportunities to enhance signal reception, mitigate channel depolarization, and suppress interference.
Building on this concept, Ding et al. \cite{ding2026polarforming} investigated the performance gains achievable through dynamic polarization control and demonstrated its advantages over conventional fixed-polarization configurations. In \cite{mehrzad2026polarforming}, a joint polarforming scheme integrating a flexible intelligent metasurface (FIM) and an intelligent reconfigurable surface (IRS) was proposed to mitigate channel depolarization and maximize the achievable sum-rate using a generative adversarial network (GAN)-assisted meta-soft actor-critic (Meta-SAC) framework. Furthermore, Zhou et al. \cite{zhou2025polarforming} incorporated polarforming into movable-antenna systems, jointly optimizing antenna positions and polarization configurations through a successive convex approximation (SCA)-based algorithm to maximize the achievable rate.
In \cite{shao2025polarforming}, the authors extended polarforming to integrated sensing and communication (ISAC) systems by jointly optimizing electrical polarforming and mechanical antenna configurations to maximize the sum-rate, demonstrating significant gains over conventional systems.

Despite these advances, the application of polarforming to SWIPT systems remains unexplored. To address this gap, this work introduces dynamic polarization control at both the base station (BS) and user devices in a SWIPT network, where each user employs power-splitting hardware to simultaneously perform information decoding (ID) and energy harvesting (EH). We formulate a non-convex sum-rate maximization problem subject to minimum instantaneous harvested power constraints. To handle the large number of coupled variables, we adopt a GNN-based learning approach in which network devices serve as nodes and are updated iteratively using channel state information (CSI). Although our model considers instantaneous harvested power, we use the conventional term energy harvesting to maintain consistency with the SWIPT literature. The work offers three main contributions.
\begin{itemize}
	\item This paper proposes a downlink system in which a BS and multiple users are both equipped with dynamic polarization (polarforming). Each user also supports simultaneous information decoding and energy harvesting via power-splitting RF hardware.
	\item We adopt an imperfect channel model to ensure that the analysis of our system more closely reflects realistic propagation conditions. The numerical results section offers a performance comparison between the proposed dynamic polarization scheme and the baseline fixed-polarization schemes (linear and circular polarization).
	\item To address the non-convex optimization problem, we employ a GNN-based framework over the considered network graph. In the numerical results section, the performance of the proposed GNN approach is compared against a conventional DNN-based baseline.
\end{itemize}
In this paper, the sections are organized as follows: First, we present the system and signal model in Section \ref{System Model and Opt Prob}. Section \ref{Alg-sec} describes the proposed solution. The simulation results are then analyzed in Section \ref{Simulation}. Finally, Section \ref{Conclusion} summarizes the paper.
\vspace{-0.7em}
\section{System and Signal Model} \label{System Model and Opt Prob}
\subsection{System Model}
We propose a downlink multi-user system in which the BS is equipped with a uniform rectangular array (URA) consisting of \(N = N_x N_z\) elements, where \(N_x\) and \(N_z\) denote the number of elements along the \(x\)- and \(z\)-axes, respectively. The position of the \(n\)-th URA element is given by \(\boldsymbol{\kappa}_n = [x_n, \, y_n, \, z_n]^T \in \mathbb{R}^{3}\). The BS serves \(K\) users, each equipped with a single dual-polarized antenna, that simultaneously functions as both an energy harvesting receiver (EHR) and an information decoding receiver (IDR). We define the index sets \(\mathcal{N} = \{1, 2, \dots, N\}\) and \(\mathcal{K} = \{1, 2, \dots, K\}\) for the antennas and users, respectively.
Each antenna element at both the BS and the users consists of two orthogonally polarized ports, vertical and horizontal, along with a phase shifter that dynamically adjusts the phase of each port \cite{mehrzad2026polarforming}. To model this behavior, the polarforming vector (PFV) for the \(n\)-th BS antenna element is defined as
\vspace{-0.3em}
\begin{align}
	\mathbf{e}_n = \frac{1}{\sqrt{2}} \, \left [e^{ \jmath \theta_{1,n}}, \, e^{ \jmath  \theta_{2,n}} \right ]^T, \quad \forall n \in \mathcal{N},
\vspace{-0.3em}	
\end{align}
where $\theta_{1,n}, \, \theta_{2,n} \in [0, \, 2\pi)$ are the phase shifts of the $n$-th antenna element at the BS, and $\frac{1}{\sqrt{2}}$ ensures normalization to comply with the transmit power constraint. Similarly, the receive PFV at user $k$ is given by $\mathbf{q}_k = \left [e^{\jmath \phi_{1,k}}, \, e^{\jmath \phi_{2,k}} \right ]^T,\forall k \in \mathcal{K}$, with $\phi_{1,k}, \, \phi_{2,k} \in [0, \, 2\pi)$ representing the corresponding phase shifts.
\begin{figure}[t!]
	\begin{center}
		\includegraphics[scale = 0.08, trim={5 2 5 5},clip]{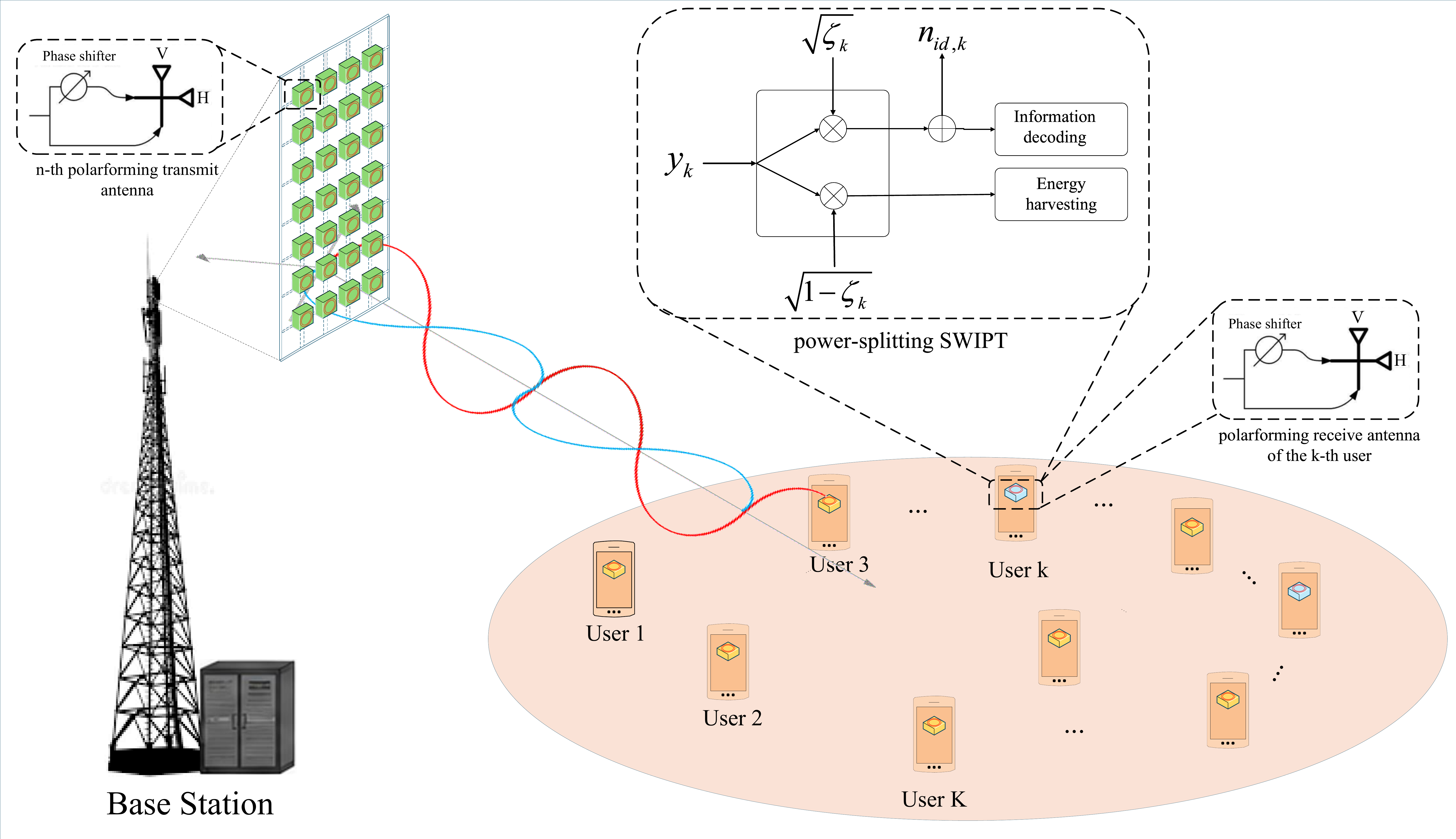}
		\caption{Polarforming-enabled SWIPT system model.}%
		\label{fi:system}%
		\vspace{-13pt}
	\end{center}
	\vspace{-0.3em}	
\end{figure}

For mathematical conventions, polarforming matrix (PFM) at the BS is defined as
\begin{align}
	& \mathbf{E} (\boldsymbol{\theta})= \textrm{blkdiag} \left \{ \mathbf{e}_1, \, \mathbf{e}_2, \, \dots , \, \mathbf{e}_N\right \} \in \mathbb{C}^{2N \times N}.
\end{align}
In this letter, we adopt a power-splitting (PS) SWIPT architecture at the users. Each user splits the received power into two separate streams: 1) information decoding (ID), and 2) energy harvesting (EH). For this purpose, the user $k$ employs a PS ratio $\zeta_k \in [0, 1], \forall k \in \mathcal{K}$. During the ID process, additional noise is introduced by the receiver circuitry. This noise is modeled as complex additive white Gaussian noise (AWGN), where represents $n_{id,k} \sim \mathcal{CN} (0, \sigma^2_{id, k}), \forall k \in \mathcal{K}$. 
\vspace{-0.5em}	 
\subsection{Channel Model}
We adopt a Rician fading channel model with quasi-static far-field flat-fading between the BS and the users. The channel consists of two components: 1) LOS, and 2) NLOS. The steering vector is denoted by $\mathbf{a}_{k}  \, (\vartheta_k, \varphi_k) \in \mathbb{C}^{N \times 1}, \forall k \in \mathcal{K}$, and is expressed as:
\vspace{-0.5em}	
\begin{align}
	\mathbf{a} \, (\vartheta, \varphi) = \left [ e^{j \frac{2\pi}{\lambda} \mathbf{c}^T \boldsymbol{\kappa}_1}, \, e^{j \frac{2\pi}{\lambda} \mathbf{c}^T \boldsymbol{\kappa}_2}, \dots, \, e^{j \frac{2\pi}{\lambda} \mathbf{c}^T \boldsymbol{\kappa}_N}\right ]^T,
\vspace{-0.5em}		
\end{align}
where $\vartheta \in [-\frac{\pi}{2},  \frac{\pi}{2}]$ and $\varphi \in [0, \, \pi]$ denote the azimuth and elevation angles of departure (AoD), respectively, and $ \lambda$ is the wavelength. The vector $\mathbf{c} \triangleq [ \, \sin (\varphi) \cos (\vartheta), \, \sin (\varphi) \sin (\vartheta), \, \cos(\varphi)]^T$ represents the propagation direction. Accordingly, the Rician polarized channel from the BS to the $k$-th user is given by \cite{li2026polarforming}:
\vspace{-0.5em}	
\begin{align}
	\mathbf{G}_k = \sqrt{\frac{\beta \mu_{k}  }{1+\beta}} \underbrace{\mathbf{a}_{k} \otimes \boldsymbol{\Lambda}}_{ \text{\tiny LOS}}  + \sqrt{\frac{\mu_{k} }{1+\beta}} \underbrace{\tilde{\mathbf{g}}_k \,  \otimes \boldsymbol{\Lambda}}_{\text{\tiny NLOS}}, 
\vspace{-0.5em}		
\end{align}
where $\mathbf{G}_{k}\in \mathbb{C}^{2N \times 2}$. The parameters $\mu_k$ and $\beta$ denote the path loss between the BS and $k$-th user and the Rician factor, respectively. The matrix $\boldsymbol{\Lambda} \in \mathbb{R}^{2 \times 2}$ represents the channel polarization matrix. To account for non-ideal isolation between the co- and cross-polarization components at the receiver antenna, $\boldsymbol{\Lambda}$ is modeled as:
\vspace{-0.5em}	
\begin{align}
	\boldsymbol{\Lambda}  = & \frac{1}{\sqrt{1 + \upsilon}}
	\begin{bmatrix}
		1 & \sqrt{\upsilon} \\
		\sqrt{\upsilon} & 1 
	\end{bmatrix},
	\vspace{-0.5em}	
\end{align}
where $\upsilon$ represents the antenna inverse cross-polarization discrimination (XPD) parameter. Additionally, $[\tilde{\mathbf{g}}_k]_i$ is assumed to follow $ \mathcal{CN} (0, 1)$. Finally, the effective channel from the BS to user $k$ is given by
\vspace{-0.5em}	
\begin{align}
	\mathbf{h}_k = \mathbf{E}^H (\boldsymbol{\theta})\, \mathbf{G}_k \, \mathbf{q}_k \quad\in \mathbb{C}^{N\times 1}.	
\vspace{-0.5em}		
\end{align}
To reflect a realistic scenario, we assume that perfect CSI is unavailable. Moreover, CSI imperfections complicate polarization state matching and compensation. Accordingly, the channel is modeled as $\widehat{\mathbf{h}}_k = \mathbf{h}_k + \Delta \mathbf{h}_k$, where $\Delta \mathbf{h}_k \sim \mathcal{CN} \, (\mathbf{0}, \xi \mathbf{I}_N)$ is the channel estimation error and $\xi$ is the error variance.
\vspace{-0.8em}	
\subsection{Signal Model}
The BS transmits an independent and identically distributed (i.i.d.) information symbol $s_k \in \mathbb{C}$ with the zero mean and unit variance to the user $k$. The transmitted signal can be written as $\mathbf{x} =  \textstyle \sum_{k \in \mathcal{K}} \mathbf{b}_k s_k$, where $\mathbf{b}_k \in \mathbb{C}^{N \times 1}$ represents the digital beamforming vector. The received signal at user $k$ is then $y_k = \widehat{\mathbf{h}}_{k}^H \mathbf{x} + n_k$, where $n_k$ is the AWGN at the user $k$ with zero mean and variance $\sigma_k^2$. At the user $k$, the received power is split into two parts using the PS factor \(\zeta_k\). The resulting signals for ID and EH are respectively given by
\begin{align}
	y_{ID, k} & =  \sqrt{\zeta_k} \, y_k + n_{id,k},\\ 
	y_{EH, k} & = \sqrt{1 - \zeta_k} \, y_k.
\end{align}
The achievable rate at the user $k$ can be calculated as $\mathcal{R}_k = \log_2 (1 + \gamma_k)$, where $\gamma_k = \frac{\zeta_k \, \left \lvert \widehat{\mathbf{h}}_k^H \, \mathbf{b}_k \right \rvert^2}{\zeta_k \, \textstyle \sum_{{k^{\prime}} \in \mathcal{K}, {k^{\prime}} \ne k} \left \lvert \widehat{\mathbf{h}}_{k}^H \, \mathbf{b}_{k^{\prime}} \right \rvert^2 + \zeta_k \sigma_k^2 + \sigma_{id,k}^2}$ represents the signal-to-interference-plus-noise ratio (SINR). The EH by the $k$-th user is obtained as $P_{k} = \chi \, (1-\zeta_k) \, \left (\textstyle \sum_{k^{\prime} \in \mathcal{K}} \left  \lvert  \widehat{\mathbf{h}}_k^H \, \mathbf{b}_{k^{\prime}} \right \rvert^2 + \sigma_k^2 \right )$, where $\chi$ denotes energy conversion efficiency factor. 
\vspace{-1em}	
\subsection{Optimization Problem}
Our objective is to maximize the achievable sum-rate of all users by jointly optimizing the digital beamforming vectors, the BS PFVs, the receive PFVs, and the PS ratios. The optimization problem is formulated as follows:
\begin{subequations}
	\small
	\begin{align}
		\mathcal{P}_{1}:\underset{{\{\mathbf{b}_{k}\}, \{\mathbf{e}_{n}\}, \{\mathbf{q}_{k}\}, \{\zeta_{k}\}}}{\text{maximize}}  \quad &  \textstyle\sum_{k \in \mathcal{K}} \mathcal{R}_{k}  \notag \\
		\text{s.t.} \quad & \text{C}_1:  \mathcal{R}_{k} \ge R_{th}, \quad k \in \mathcal{K}, \notag\\
		\quad  & \text{C}_2: P_t \leq P_{\text{BS}}^{\max},  \notag\\
		\quad & \text{C}_3:  P_k \ge P_{th}, \quad k \in \mathcal{K},\notag\\
		\quad & \text{C}_4:  \theta_{1,n}, \theta_{2,n} \in[0,2 \pi), \quad n \in \mathcal{N},\notag\\
		\quad & \text{C}_5:  \phi_{1,k}, \phi_{2,k} \in[0,2 \pi), \quad k \in \mathcal{K},\notag\\
		\quad & \text{C}_6:  \zeta_k \in [0, 1], \quad k \in \mathcal{K}.  \notag
	\end{align}
	\label{opt_prob}
\end{subequations}
Constraint $\text{C}_1$ ensures a minimum achievable rate for each user. Constraint $\text{C}_2$ limits the total transmit power $P_t =  \textstyle\sum_{k \in \mathcal{K}} \|\mathbf{b}_k\|^2$ to the maximum BS transmit power $P_{BS}^{\max}$. Constraint $\text{C}_3$ enforces a minimum harvested power for each user. Constraints $\text{C}_4$ and $\text{C}_5$ restrict the phase shifts of the polarized antenna elements at the BS and the users, respectively. Finally, constraint $\text{C}_6$ bounds the PS ratio within feasible range. The problem $\mathcal{P}_{1}$ is non-convex and involves highly coupled variables. To address the limitations associated with conventional optimization methods, we propose a learning-based approach using a graph neural network (GNN). Specifically, we employ a readout layer at the output layer to enforce certain constraints by construction, while a penalty term is added to the loss to encourage satisfaction of the remaining constraints during training.  
\section{The Proposed Graph Neural Network Framework} \label{Alg-sec}
In this section, we introduce the proposed GNN and its components, and explain its operational mechanism. We then develop a graph representation for our system. Finally, we design a custom loss function to optimize the proposed optimization problem.
\vspace{-1em}	
\subsection{Architecture of the Proposed GNN}
The proposed system comprises $K+1$ heterogeneous nodes, of which $K$ represent users and the remaining one denotes the BS. At each user node $k$, the associated optimization variables, including the digital beamforming vectors, PS ratios, and receive PFVs, are updated simultaneously across all nodes. Meanwhile, the PFM is derived at the BS node. The nodes are linked via edges, and bidirectional communication is assumed between every pair of nodes $i$ and $j$ for information exchange. Furthermore, we assume that every user node can share its local information with the BS. The node operations and information exchange are presented in three layers, as detailed below:
\subsubsection{Initial Layer}
In the initial layer, the raw parameters of the proposed system are first mapped to a feature vector set denoted as $\mathcal{V}^{(0)}$, where  $[\mathbf{v}_{1}^{(0)}, \dots, \mathbf{v}_{K}^{(0)}, \mathbf{v}_{bs}^{(0)}] \in \mathcal{V}^{(0)}$, enabling the data to be prepared for subsequent processing stages and message passing among nodes. The output of the initial layer for user node $k$ is then computed by concatenating the real and imaginary parts of its CSI vector as the input as $\mathbf{v}_k^{(0)} = T_{u}^{(0)} \, ([\Re\{\widehat{\mathbf{h}}_k\}, \Im\{\widehat{\mathbf{h}}_k\}]) \in \mathbb{R}^{w} $, where $T_{u}^{(0)}(\cdot)$ represents the feature extraction function for user nodes, and $w$ is a configurable feature dimension parameter. Similarly, for the BS node, the output vector is obtained as $\mathbf{v}_{bs}^{(0)} = T_{bs}^{(0)}([\underset{k}{\textrm{mean}}(\Re\{\widehat{\mathbf{h}}_k\}),\underset{k}{\textrm{mean}}(\Im\{\widehat{\mathbf{h}}_k\})])   \in \mathbb{R}^{w}$, where $T_{bs}^{(0)}(\cdot)$ and $\underset{i}{\textrm{mean}} (x_i)$ are the corresponding feature extraction function for the BS node and the element-wise mean over all $x_i$, respectively.
\subsubsection{Node Messaging and Update Layer}
After extracting the initial features, a multi-layer structure is considered to pass messages between different nodes and share information. In such a way that we denote the output feature vector of the user node $k$ by $\mathbf{v}_k^{(\ell)}$ and the node associated with the BS by $\mathbf{v}_{bs}^{(\ell)}$ in the layer $\ell$. Additionally, based on the edges between users as well as between users and the BS, we perform edge feature extraction and node update for the subsequent layer. Accordingly, the features of all inter-node edges are extracted using the following feature extraction functions:
\begin{align}
	&\mathbf{r}^{(\ell)}_{k} = T_{edge,bu}^{(\ell)} ([\mathbf{v}_{bs}^{(\ell-1)}, \mathbf{v}_k^{(\ell-1)} ]) \in \mathbb{R}^w, \quad k \in \mathcal{K}, \\
	&\mathbf{r}^{(\ell)}_{i,j} = T_{edge,uu}^{(\ell)} ([\mathbf{v}_i^{(\ell-1)}, \mathbf{v}_j^{(\ell-1)}]) \in \mathbb{R}^w, \quad (i,j) \in \mathcal{K}^2,
\end{align}
where $ T_{edge,bu}^{(\ell)}(\cdot)$ and $T_{edge,uu}^{(\ell)}(\cdot)$ denote the feature extraction functions for BS-to-user and user-to-user edges at layer $\ell$, respectively. Here, $\ell \in \{1, \dots, L\}$, and the model comprises $L$ layers in total.

For node updating, the edge features and the features from layer $\ell -1$ are aggregated. The following functions are then applied:
\begin{align}
	&\mathbf{v}_k^{(\ell)} = \left [ T_{u}^{(\ell)} \, \left ([\mathbf{v}_k^{(\ell-1)}, \mathbf{r}^{(\ell)}_{k}, \frac{1}{K-1}\sum_{j \ne k}\mathbf{r}^{(\ell)}_{k,j}] \right ),\mathbf{v}_k^{(\ell-1)}\right ], \\
	&\mathbf{v}_{bs}^{(\ell)} = \left [ T_{bs}^{(\ell)} \, \left (\left[\mathbf{v}_k^{(\ell-1)}, \underset{k}{\mathrm{mean}}  (\mathbf{r}^{(\ell)}_{k} )\right]\right ),\mathbf{v}_{bs}^{(\ell-1)}\right ],
\end{align}
where $T_{u}^{(\ell)} (\cdot)$ and $T_{bs}^{(\ell)} (\cdot)$ represent the node update functions for user nodes and the BS nodes, respectively. In each update step, the edge features and the features from layer $\ell -1$ are aggregated to compute the node representations for layer $\ell$. Consequently, the dimension of the node feature vectors grows with each layer, such that $\mathbf{v}_k^{(\ell)}, \mathbf{v}_{bs}^{(\ell)} \in \mathbb{R}^{ 2 \ell w}$ for user nodes and the BS node. 
\begin{figure}[t]
	\centering
	\includegraphics[height=4.5cm, keepaspectratio]{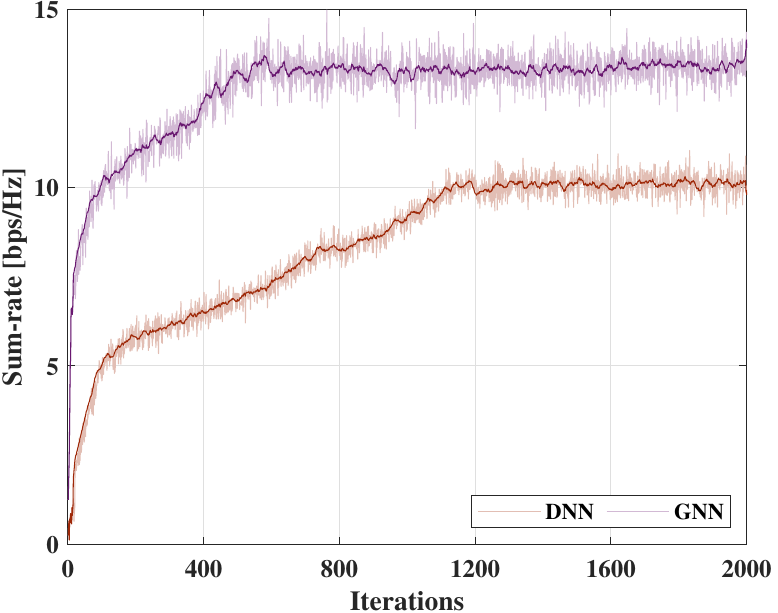}
	\caption{Convergence performance comparison of the proposed GNN framework and DNN.}
	\label{fig1}
	\vspace{-7pt}
\end{figure}
\subsubsection{Readout Layer}
After $L$ layers of updates, the GNN produces  a candidate solution to problem $\mathcal{P}_1$. Accordingly, the readout layer is designed based on the input feature vectors of the nodes, namely $\mathbf{v}_k^{(L)}, \mathbf{v}_{bs}^{(L)} \in \mathbb{R}^{(L+1)\,w}$ for user nodes and the BS node. The optimization variables are then derived as follows:
\begin{itemize}
	\item \textbf{The digital beamforming vector}: For each user $k$, we compute $\mathbf{b}_k^{(nn)} = T_{b}(\mathbf{v}_k^{(L)}) \in \mathbb{R}^{2N}$. The complex digital beamforming vector is then constructed as $\bar{\mathbf{b}}_k = [b_{k,1}^{(nn)}, \dots, b_{k,N}^{(nn)}]^T + \jmath \cdot \, [b_{k,N+1}^{(nn)}, \dots, b_{k,2N}^{(nn)}]^T$. To satisfy constraint $\text{C}_2$, the beamforming vector is normalized as $\mathbf{b}_k = \sqrt{P_{BS}^{\max}/\sum_k \lVert \bar{\mathbf{b}}_k \rVert^2} \, \bar{\mathbf{b}}_k$. Notably, in a sum-rate maximization system, Shannon's formula suggests that using the maximum available power benefits both the ID and EH of each user.
	\item \textbf{The PS ratio}:
	The PS ratio for user $k$ is obtained as $\zeta_k^{(nn)} = T_{\zeta}(\mathbf{v}_k^{(L)}) \in \mathbb{R}$. To enforce constraint $\text{C}_6$, we apply the sigmoid function as $\zeta_{k} = \mathrm{Sigmoid} (\zeta_k^{(nn)})$.
	\item \textbf{The receive PFV}: Using the mapping function $T_{\phi}(\cdot) \in \mathbb{R}^2$, the input $\mathbf{v}_k^{(L)}$ outputs $\boldsymbol{\phi}_k^{(nn)} = [\phi_{1,k}^{(nn)}, \phi_{2,k}^{(nn)}]$. To satisfy constraint $\text{C}_5$, the vector is scaled as $\boldsymbol{\phi}_k = 2 \pi \, \mathrm{Sigmoid}\, (\boldsymbol{\phi}_k^{(nn)})$. The receive PFV $\mathbf{q}_k$ can then be obtained accordingly.
	\item \textbf{The PFM}: To obtain the PFM at the BS, we compute $\boldsymbol{\theta}^{(nn)} = T_{\theta} (\mathbf{v}_{bs}^{(L)}) \in \mathbb{R}^{2N}$, where $\boldsymbol{\theta}^{(nn)} = [\theta^{(nn)}_{1,1}, \theta^{(nn)}_{1,2}, \dots, \theta^{(nn)}_{N,1}, \theta^{(nn)}_{N,2}]$. The scaled vector is given by $\boldsymbol{\theta} = 2 \pi \, \mathrm{Sigmoid} \, (\boldsymbol{\theta}^{(nn)})$, from which the PFM, $\mathbf{E} (\boldsymbol{\theta})$ can be computed.   
\end{itemize}

Given the discussions on constraints in the preceding subsections, we design a custom loss function to penalize violations of the remaining constraints during training of the entire proposed model. Accordingly, the proposed loss function is formulated as follows:
\begin{align}
	\mathcal{L} = - \sum_{k \in \mathcal{K}} \mathcal{R}_k &+ \sum_{k \in \mathcal{K}} \rho_k \, \mathrm{ReLU} (R_{th} - \mathcal{R}_k) \notag \\ &+ \sum_{k \in \mathcal{K}} \eta_k \, \mathrm{ReLU} (P_{th} - P_k),
\end{align}
\begin{table}[t]
	\centering
	\caption{The proposed system model and GNN framework parameters}
	\label{system_parameters}
	\begin{tabular}{|c | c | c |c|} 
		\hline
		\textbf{Parameter} & \textbf{Value} & \textbf{Parameter} & \textbf{Value} \\
		\hline
		
		$P_{BS}^{\max}$        & $30 \, \text{dB}_m$        & $N$        & $16$        \\ \hline
		$R_{th}$        &  $0.4 \, \text{bps/Hz}$        &  $K$        &  $4$        \\ \hline
		$P_{th}$        & $-40 \, \text{dB}_m$        &      $\beta$ & $0 \, \text{dB}$   \\ \hline
		$\sigma_k^2 = \sigma_{id,k}^2$        & $-60 \, \text{dB}_m$  & Learning rate        & $0.001$              \\
		\hline
		$\nu$ & $0.3$ &  Batch size   & $32$ \\ \hline 
		Frequency & $6 \, \text{GHz}$ &   L & $3$ \\ \hline 
		Antenna spacing & $\frac{\lambda}{2}$ &  Number of iterations  & $2000$ \\ \hline
	\end{tabular}
\end{table} 
where $\rho_k$ and $\eta_k$ are positive penalty coefficients. The terms $\mathrm{ReLU} (R_{th} - \mathcal{R}_k)$ and $\mathrm{ReLU} (P_{th} - P_k)$ impose penalties when constraints $\text{C}_1$ and $\text{C}_3$ are violated, respectively.
\vspace{-1em}
\section{Numerical Results} \label{Simulation}
We present the simulation environment of the proposed system model, where the BS is located at $(0, 0, 0)$ meters and the users are uniformly distributed at random locations within a 3D box defined by $x_k \in [-20, +20]$, $y_k \in [0, +20]$, and $z_k \in [0, -20]$. In the proposed GNN-based framework, each feature extractor $T_i(\cdot)$ is implemented as a two-layer fully connected network with $128$ neurons per layer and an output dimension of $w = 128$. The baseline DNN employs a three-layer fully connected architecture, also with 128 neurons in each layer. The path loss $\mu_k$ for user $k$ is obtained as  $\mu_0 \, d_k ^ {-\psi}$, where $\mu_0 = (\frac{\lambda}{4 \pi})^2$, $\psi = 3$ is the path loss exponent, and $d_k$ is the distance from the BS to user $k$. The remaining simulation parameters are summarized in Table \ref{system_parameters}.

\begin{figure*}[t]
	\centering
	\begin{subfigure}[b]{0.20\textwidth} 
		\includegraphics[
		width=4.5cm,
		height=3.5cm,
		trim={0 1 1 1},
		clip
		]{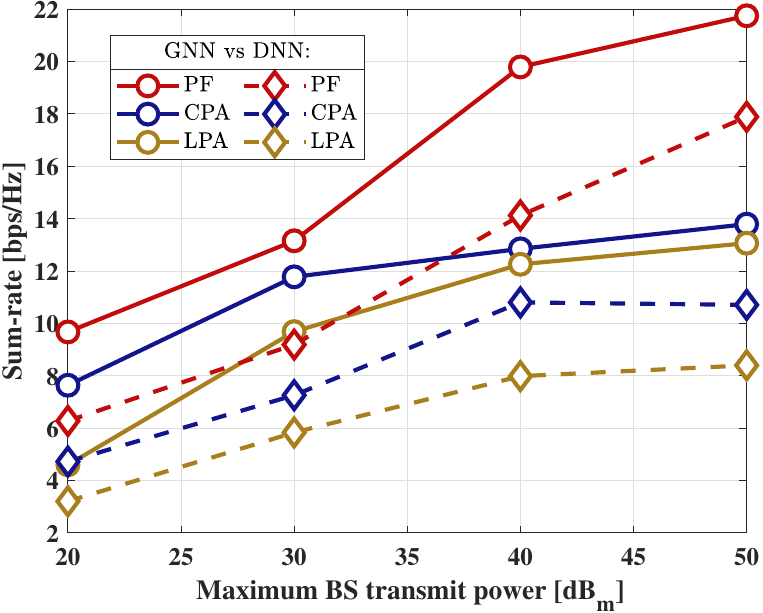}
		\caption{Sum-rate vs. the maximum BS transmit power.}
		\label{fig2}
	\end{subfigure}
	\hspace{0.04\textwidth} 
	\begin{subfigure}[b]{0.20\textwidth}
		\includegraphics[
		width=4.5cm,
		height=3.5cm,
		trim={0 1 1 1},
		clip
		]{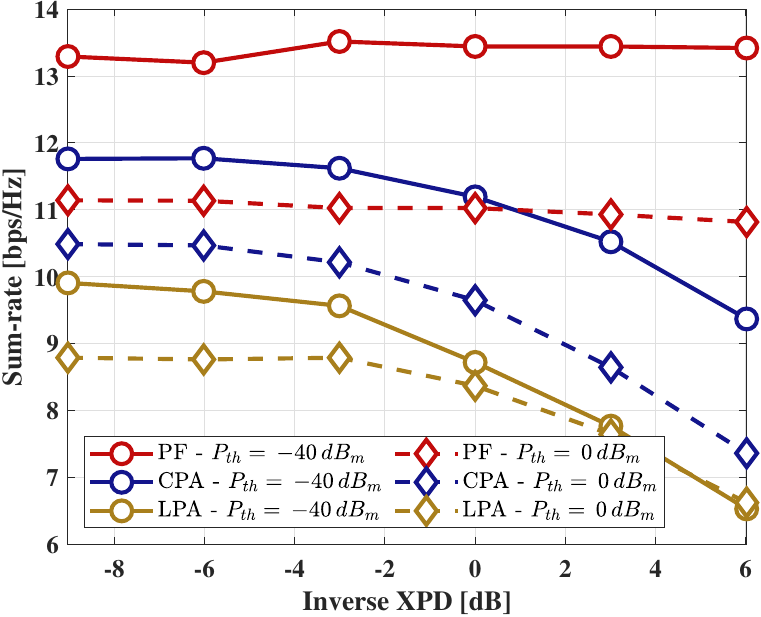}
		\caption{Sum-rate vs. the inverse XPD.}
		\label{fig3}
	\end{subfigure}
	\hspace{0.04\textwidth}
	\begin{subfigure}[b]{0.20\textwidth}
		\includegraphics[
		width=4.5cm,
		height=3.5cm,
		trim={0 1 1 1},
		clip
		]{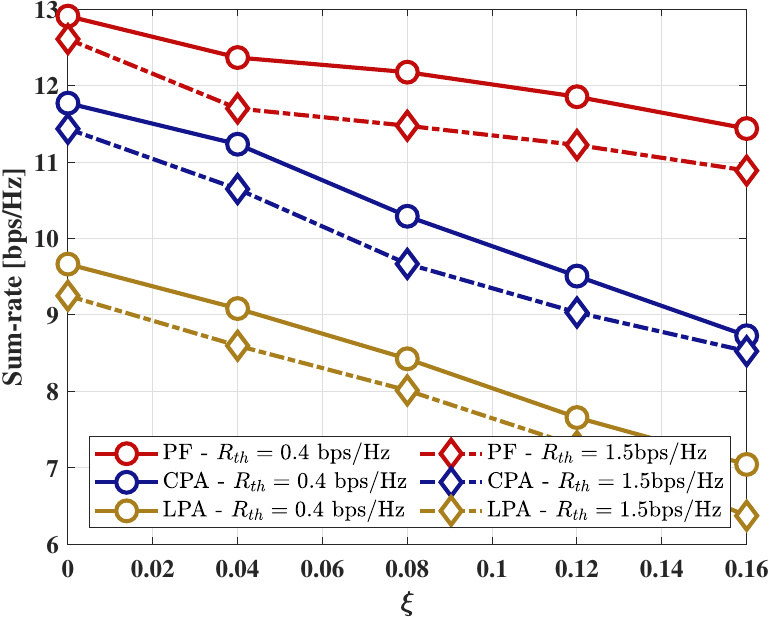}
		\caption{Sum-rate vs. the CSI error variance.}
		\label{fig4}
	\end{subfigure}
	\hspace{0.04\textwidth}
	\begin{subfigure}[b]{0.20\textwidth}
		\includegraphics[
		width=4.5cm,
		height=3.5cm,
		trim={0 1 1 1},
		clip
		]{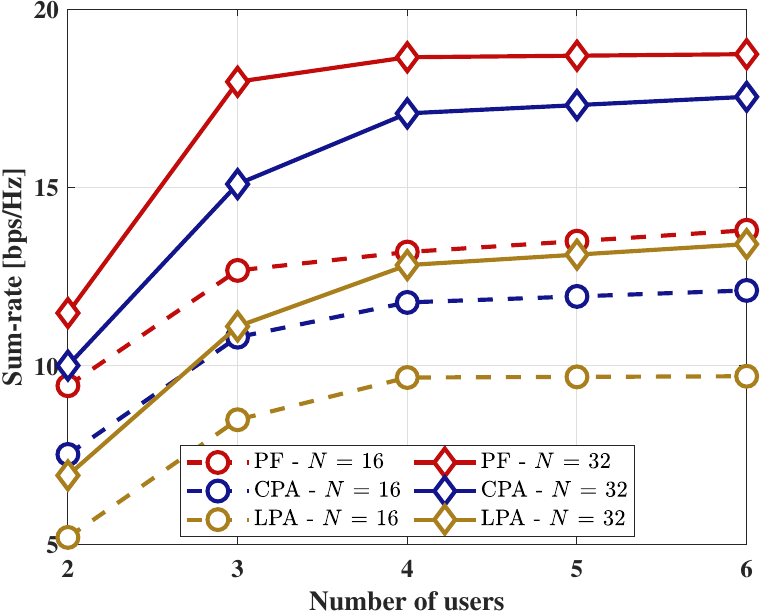}
		\caption{Sum-rate vs. the number of users.}
		\label{fig5}
	\end{subfigure}	
	\caption{Simulation results.}
	\label{Fig0}
	\vspace{-1.3em}
\end{figure*}

Fig. \ref{fig1} depicts the convergence behavior of the proposed GNN framework compared with the DNN in terms of achievable sum-rate. The results demonstrate the enhanced performance of the GNN framework, which achieves both faster and higher convergence than the DNN. In particular, the proposed GNN framework converges within approximately 400 iterations, whereas the DNN requires around 1200 iterations to stabilize.  
Fig. \ref{fig2} illustrates the impact of the BS transmit power on the achievable sum-rate. The proposed scheme, which employs a polarforming (PF) via the GNN framework, achieves improved performance compared to the DNN-based optimization. For further comparison, we also implement benchmarks based on a linearly polarized antenna (LPA) and a circularly polarized antenna (CPA) as fixed-polarization schemes. The PF with the GNN framework provides an average achievable sum-rate improvement of $39.7 \%$ over the DNN-based approach. Under the same GNN framework, the PF achieves average gains of $37.5 \%$ and $68.5 \%$ against the CPA and LPA benchmarks, respectively.

Fig. \ref{fig3} shows the achievable sum-rate versus the inverse XPD. In fact, an increase in the inverse XPD corresponds to stronger channel depolarization. As the inverse XPD increases, a well-designed scheme should remain non-decreasing in sum-rate. The proposed PF scheme, for two EH thresholds $P_{th} \in \{-40, 0\}\, \text{dB}_m$, mitigates the performance degradation caused by channel depolarization. In contrast, for the LPA and CPA benchmarks, increasing the inverse XPD leads to a noticeable decrease in the achievable sum-rate.

Fig. \ref{fig4} depicts the achievable sum-rate versus the imperfect CSI error variance $\xi$, which is crucial for assessing performance in practical scenarios. As expected, greater channel uncertainty reduces the sum-rate. In this figure, we also analyze the impact of the quality of service (QoS) constraint by considering minimum data rate thresholds of $R_{th} \in \{0.4, 1.5\}$ bps/Hz. Increasing $R_{th}$ imposes a stricter minimum rate requirement on every user, irrespective of channel quality. As a result, the BS is forced to prioritize resource allocation toward users with weaker channels, reducing the achievable sum-rate.
The achievable sum-rate versus the number of users is shown in Fig. \ref{fig5}. The sum-rate increases with the number of users, as the system can effectively manage inter-user interference through the available spatial resources. As expected, increasing the number of BS antennas also leads to a higher sum-rate. The proposed PF scheme consistently outperforms both the CPA and LPA. For the $N = 16$ scenario, the polarforming achieves average improvements of $16.4 \%$ and $49.8 \%$ over the CPA and LPA, respectively.

\section{Conclusion} \label{Conclusion}
In this paper, we investigated the use of polarforming in a SWIPT system. By equipping both the base station and users with the dynamically controlled antenna polarization, we leveraged an additional degree of freedom to enhance system performance. We formulated an optimization problem to maximize the achievable sum-rate, subject to constraints on power, minimum rate, and harvested energy. To solve it, we developed a GNN-based framework to learn the sub-optimal variables. Our simulation results demonstrate that the proposed polarforming scheme, trained via the GNN, consistently outperforms fixed-polarization baselines and a conventional DNN. Furthermore, we analyzed a realistic scenario involving imperfect CSI and its impact on the achievable sum-rate. The system proved robust against depolarization effects, highlighting the potential of combining polarforming with GNN-based optimization for SWIPT systems.

\bibliographystyle{IEEEtran}
{\footnotesize
	\bibliography{library}}

%
%

\vfill
 
\end{document}